\documentclass[journal, 10pt,twocolumn]{IEEEtran}
 
\usepackage{caption}
\usepackage{subcaption}
\usepackage{graphicx}
\usepackage{color,soul}
\usepackage{amsthm}
\theoremstyle{plain}

\theoremstyle{definition}

\theoremstyle{remark}

\usepackage{cite}
\usepackage{amsmath,amssymb,amsfonts}
\usepackage{algorithm}
\usepackage{algorithmic}
\usepackage{graphicx}
\usepackage{textcomp}
\usepackage{xcolor} 
\usepackage{xpatch}
\makeatletter
\ExplSyntaxOn
\cs_new:Npn \bibColoredItems #1#2
  {
    \clist_map_inline:nn {#2} { \cs_new:cpn {bib@colored@##1} {#1} } 
  }
\ExplSyntaxOff

\newcommand\bib@setcolor[1]{%
  \ifcsname bib@colored@#1\endcsname
    \expanded{\noexpand\color{\csname bib@colored@#1\endcsname}}%
  \else
    \normalcolor
  \fi
}

\IfPackageLoadedTF{hyperref}{\@tempswatrue}{\@tempswafalse}
\if@tempswa
  \xpatchcmd\@bibitem {\H@item}{\bib@setcolor{#1}\H@item}{}{\PatchFailed}
  \xpatchcmd\@lbibitem{\H@item}{\bib@setcolor{#2}\H@item}{}{\PatchFailed}
\else
  \xpatchcmd\@bibitem {\item}  {\bib@setcolor{#1}\item}  {}{\PatchFailed}
  \xpatchcmd\@lbibitem{\item}  {\bib@setcolor{#2}\item}  {}{\PatchFailed}
\fi
\makeatother

\usepackage{color}
\usepackage{bm}
\usepackage{booktabs}
\usepackage{cleveref}

\ifCLASSINFOpdf

\else

\fi

\begin{document}
\title{
\color{black}Exploiting Structured Sparsity in Near Field: \\
From the Perspective of Decomposition
}

\author{Xufeng~Guo,~Yuanbin Chen,~Ying~Wang,~\IEEEmembership{Member,~IEEE},~and~Chau~Yuen,~\IEEEmembership{Fellow,~IEEE}%
\vspace{-0.5cm}

  \thanks{	

    {\color{black}
    This work was supported by Beijing Natural Science Foundation under Grant 4222011,
    and in part by the BUPT Excellent Ph.D. Students Foundation under Grant CX2023145. (\textit{Corresponding author}: Ying Wang.)
    
      Xufeng Guo, Yuanbin Chen, and Ying Wang are with the State Key Laboratory of Networking and Switching Technology, Beijing University of Posts and Telecommunications, Beijing 100876, China  (e-mail:brook1711@bupt.edu.cn; chen\_yuanbin@163.com; wangying@bupt.edu.cn).
      
      Chau Yuen is with the School of Electrical and Electronics Engineering, Nanyang Technological University, Singapore 639798 (e-mail:chau.yuen@ntu.edu.sg).      
      }
  } 
}

\maketitle

\begin{abstract}

The {\color{black}structured sparsity can be leveraged in traditional far-field channels, greatly facilitating efficient sparse channel recovery by compressing the complexity of overheads to the level of the scatterer number. However, when experiencing a fundamental shift from planar-wave-based far-field modeling to spherical-wave-based near-field modeling, whether these benefits persist in the near-field regime remains an open issue.
To answer this question, this article delves into structured sparsity in the near-field realm, examining its peculiarities and challenges. In particular, we present the key features of near-field structured sparsity in contrast to the far-field counterpart, drawing from both physical and mathematical perspectives. 
Upon unmasking the theoretical bottlenecks, we resort to bypassing them by decoupling the geometric parameters of the scatterers, termed the triple parametric decomposition (TPD) framework.  It is demonstrated that our novel TPD framework can achieve robust recovery of near-field sparse channels by applying the potential structured sparsity and avoiding the curse of complexity and overhead.
}
\end{abstract}


\IEEEpeerreviewmaketitle

\section{Introduction}
{\color{black}By deploying antenna entries that significantly outnumber those used in conventional multiple-input multiple-output (MIMO) or massive MIMO (mMIMO) systems, extremely large-scale antenna arrays (ELAAs) can substantially benefit from spatial multiplexing and beamforming enhancements.
This may provide a tenfold spectral efficiency enhancement for 6G wireless communication scenarios \cite{1., 1.1, 10.3}. 
{\color{black}However, by greatly increasing the Rayleigh distance, the deployment of ELAAs also brings about a host of new near-field challenges, leading to increased complexity of channel estimation and the enlarged overhead of pilot transmission.}
{\color{black}To address these challenges, the beneficial sparsity inherent in wireless channels can be leveraged to facilitate efficient channel estimation.
This sparsity is attributed to the limited scatterer observed in the wireless propagation environment.
Therefore, the direct estimate of scatterers' geometric parameters (e.g., the angles of arrival or angles of departure (AoA/AoDs)) would significantly save pilot overheads and computational complexity compared to the entry-by-entry estimation of the whole channel.
Furthermore, wireless channels not only manifest general sparsity but also deliver structured sparsity. This means that the non-zero entries in the sparse channel representations adhere to specific, predetermined patterns~\cite{11.18.1}.
Building on this insight, there are sufficient works that have investigated the compressive sensing (CS)-based approaches by leveraging sparsity and structured sparsity in traditional far-field systems~\cite{11., 22., 12.}.
}

{\color{black}
  However, the aforementioned benefits of structured sparsity in far-field communications are not guaranteed to persist in the ELAA systems under near-field conditions.
  {\color{black}
    The fundamental shift from planar-wave-based to spherical-wave-based modeling poses new \textit{physical} peculiarities, which in turn give rise to \textit{mathematical} challenges when applying structured sparsity.}}
Regarding the \textit{physical} aspect, the phase difference across the array caused by the spherical wave is not linear with respect to (w.r.t.) the antenna indices{\color{black}, expanding} as a combination of minuscule planar waves emanating from distinct directions centered around the AoA/AoDs corresponding to the significant paths~\cite{11.}.
In this case, the propagation paths cannot be characterized as angular-domain (AD) energy impulses as is typically presented in traditional far-field channel modeling. By contrast, the energy impulses associated with significant paths are observed as AD waveforms that fluctuate around the genuine AoA/AoDs, which can be referred to as the {\it power leakage} issue~\cite{26., 29., TVT}. Furthermore, the distance between the scatterers and the array also demands careful attention in the presence of spherical wave modeling. Although a custom polar-domain (PD) representation was introduced in \cite{29.} to mitigate the strong correlation between angle and distance for more effective {\color{black}near-field} channel estimation, {\color{black}the power leakage remains}.

The inherent physical characteristics in near-field communications present \textit{mathematical} challenges when applying structured sparsity.
Owing to the power leakage, the AD energy distribution generated by a significant path does not correspond to a single non-zero element in the {\color{black}sparse channel representation}. 
Instead, it produces substantial interference at irrelevant positions, termed as {\it weak sparseness}~\cite{10.3}, leading to a high incidence of false alarms in CS-based algorithms.
Furthermore, the introduction of distances necessitates a complex probabilistic model to capture the full range of structured information.
This not only leads to i) prohibitive multiplicative computational complexity w.r.t. the dimensionalities of all three variables; ii) but also poses significant challenges in algorithmic design, given that there currently exists no probabilistic modeling specifically tailored for such a three-dimensional configuration.

In view of the discussion from both physical and mathematical perspectives, we deem that several critical issues remain to be addressed to effectively leverage the sparsities in the near-field context for facilitating efficient channel estimation. Specifically, the following questions have not been clearly answered:

\begin{itemize}
\item [\textbf{Q1:}] 
How does structured sparsity in the near-field regime contrast with that in the classical far-field scenario?

\item [\textbf{Q2:}]
What specific challenges arise when applying structured sparsity in the near-field context?

\item [\textbf{Q3:}] 
{\color{black}How can we achieve efficient sparse channel estimation in near-field communications, by leveraging structured sparsity?}

\end{itemize}

In response, we {revisit the structured sparsity} originated from the classical compressed sensing theory, and present the novel characteristics brought by the near-field communication. Our contributions can be summarized as follows: 

\begin{itemize}
  {\color{black}
  \item We first highlight the fundamental physical distinctions between far-field and near-field communications, which make the far-field sparsifying method impractical in near-field systems.
  
  \item We then unmask the mathematical challenges brought by the physical peculiarities when implementing structured sparsity in near-field systems.
  
  \item 
  We demonstrate that by adopting a strategy of parametric decoupling of sparse scatterers within the channel—specifically, through a triple parametric decomposition (TPD) framework—we can bypass the challenges above. This approach opens the door to a broad spectrum of structured sparsity applications.}

\end{itemize}








\section{Revisiting Structured Sparsity: From Far Field to Near Field}

In this section, we revisit the concept of structured sparsity inherent in traditional far-field communications. We elucidate how the traditional schemes employ the features of structured sparsity within the context of planar-wave-based modeling. Subsequently, we explore the distinctions between these methodologies when adapted for their spherical-wave-based near-field counterparts.


\subsection{Revisiting Structured Sparsity in Far Field}
\textbf{\textit{What is Structured Sparsity?}}:
Before delving into structured sparsity, it is imperative to figure out the sparsity observed in conventional {\color{black}far-field systems}. In wireless propagation environments, signals transmitted from the source to its designated receiver experience a few propagation paths. 
These paths are dependent upon the clusters present in the wireless propagation environment and are termed significant paths.
{\color{black}
    Each of the significant paths generates a response sequence with linear phase differences across the antennas, in the presence of the traditional planar-wave-based assumption.
    Therefore, each array response of the corresponding scatterer can be approximately represented as an impulse signal on the angular domain by the discrete Fourier transform (DFT). Given this, the overall channel that is a linear combination of array responses of the sparse scatterers can be pruned to a sparse one, with each non-zero entry corresponding to a scatterer on the angular domain.}
Beyond such sparsity, these non-zero entries in the sparse channel matrix may exhibit specific sparsity, e.g., structured sparsity. Explicitly, scatterers tend to be distributed in clusters, leading to a phenomenon known as clustered sparsity, where non-zero entries in the AD sparse vector are more likely to group together~\cite{11.}. This specific characteristic constitutes additional structured information that can further enhance estimation accuracy and reduce overhead. 
{\color{black}By harnessing this beneficial property, the required number of observations to achieve robust sparse channel recovery can be drastically reduced, leading to much fewer communication overheads like pilots and RF chains.} For example, in a wireless system equipped with an $N$-antenna array and $L$ propagation paths, the order of the required number of RF chains and pilots can be decreased from the Nyquist sampling rate $\mathcal{O}(N/2)$ to $\mathcal{O}(L\log(N/L))$ by using sparsity~\cite{11.18, 11.18.1}.
The CS-based schemes assisted by structured sparsity can further compress this expense to between $\mathcal{O}\left(L\right)$ and $\mathcal{O}\left( L\log(N/L) \right)$ without significant performance loss \cite{22., 12.,11., 11.18.cited.5}.


\textit{\textbf{How to Use Structured Sparsity?}}:
  To leverage the additional information offered by structured sparsity in far-field MIMO systems, state-of-the-art schemes have already introduced probabilistic methods in compressive sensing~\cite{11.18}. Specifically, an orthogonal or near-orthogonal dictionary basis is necessary to transform the channel into AD sparse representation. In far-field cases, since the channel can be directly expressed as a linear combination of planar waves, the columns of the DFT matrix serve as the dictionary basis~\cite{11.}. 
  To further characterize the prior structured information, binary supports based on Markov chains, fields, or trees can be established behind the entries of the sparse vector, referred to as hidden Markov models (HMMs) to impose the structured sparsity~\cite{11.18}. 
  {\color{black}Specifically, the values and locations of the non-zero entries in the sparse signal exhibit a unique pattern that can be formulated according to the inherent characteristics of the specific channel model. By establishing the hidden support binary pattern in the HMM, we can capture such structural information in practical channels and flexibly characterize various forms of structured sparsity~\cite{11.18}.}
  {\color{black}For example, a Markov chain model has been developed to capture the clustered paths in mmWave channels~\cite{11.}. An enhanced HMM featuring a dual structure has been proposed to address the unique structured sparsity in two-hop channels within the reconfigurable intelligent surface (RIS)-aided cascaded systems~\cite{22.}. Beyond one-layer structures, a hierarchical HMM has been constructed to capture common sparsity in uplink multi-user systems, where users share the same significant paths~\cite{12.}.}

\begin{figure*}[htbp]
	\centering
  \includegraphics[width=0.85\linewidth]{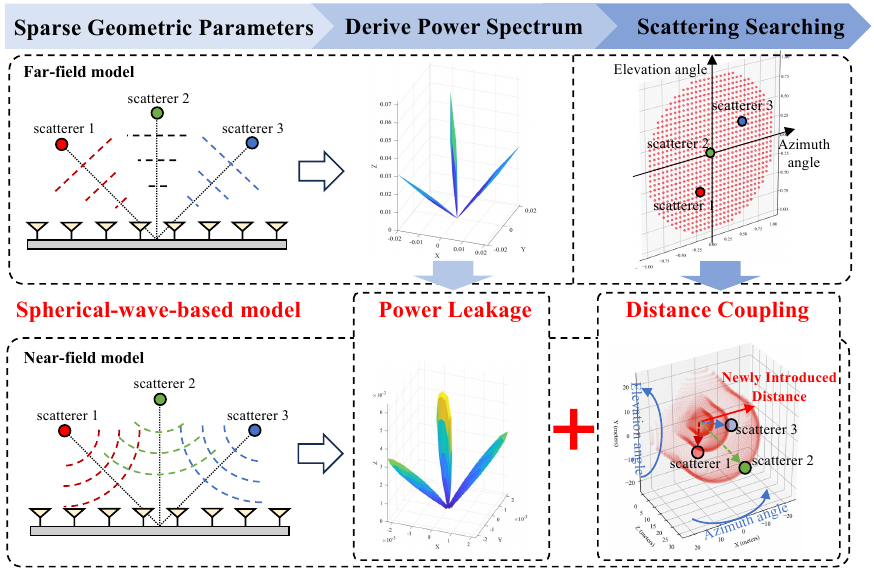}
  \caption{\color{black}
  Physical peculiarities caused by the spherical-wave-based modeling in near-field communications.}
	\vspace*{-0.5cm}
  \label{fig:Fig1}
\end{figure*}

\subsection{\color{black}
  Physical Peculiarities in Near-Field Regime}
  {\color{black}
  {\bf \textit{What Fundamentally Differs in Near Field?:}}
  In near-field communications, spherical-wave modeling becomes necessary, requiring the joint estimation of elevation and azimuth angles, as well as distances between scatterers and the base station \cite{26., 29.}. The tight coupling of these three geometric parameters poses two significant challenges for their robust recovery, i.e., the power leakage and a complex triple-coupled structure of the geometric parameters, as elaborated on in the subsections that follow.
  }
\subsubsection{\textbf{Power Leakage}}
{\color{black}
  As illustrated in Fig.~\ref{fig:Fig1}, in far-field scenarios, planar waves emanating from scatterers can be efficiently transformed into AD sparse power peaks through DFT. These peaks unambiguously identify the significant angles. 
However, this clarity is compromised in near-field scenarios. In such cases, the spherical waves generated by scatterers can be conceptually represented as a multitude of micro-planar wavefronts. These wavefronts originate from a range of directions that are centered around the original significant angles. Consequently, the AD power distributions that were initially focused in distinct power peaks begin to spread into adjacent angular directions, thereby forming diffused lobes. This occurrence is known as the {\it power leakage} issue~\cite{29.}.}

\subsubsection{\textbf{Triple-Coupled Geometric Parameters}}
    In planar-wave-based far-field scenarios, a two-dimensional (2D) AD sparse vector is sufficient to characterize the angular directions. Upon this, identifying the directions by the clear power peaks, i.e., the elevation-azimuth angle pair would provide the complete far-field channel state information. 
    However, near-field scenarios introduce additional complexities due to the spherical nature of wave propagation. 
    In this context, each spherical wave induced by the scatterer is associated not just with the elevation-azimuth angle pair but is also intricately coupled with the distance between the scatterer and the antenna array. 
    As a result, it becomes necessary to incorporate this additional degree of freedom (DoF) related to distance into the original 2D space, leading to an intricate triple-coupled structure. Such a structure not only exacerbates the computational complexity but also hampers the capture of structured information.

\begin{figure*}[t]
	\centering
	\begin{minipage}[b]{0.45\linewidth}
		\includegraphics[width=\linewidth]{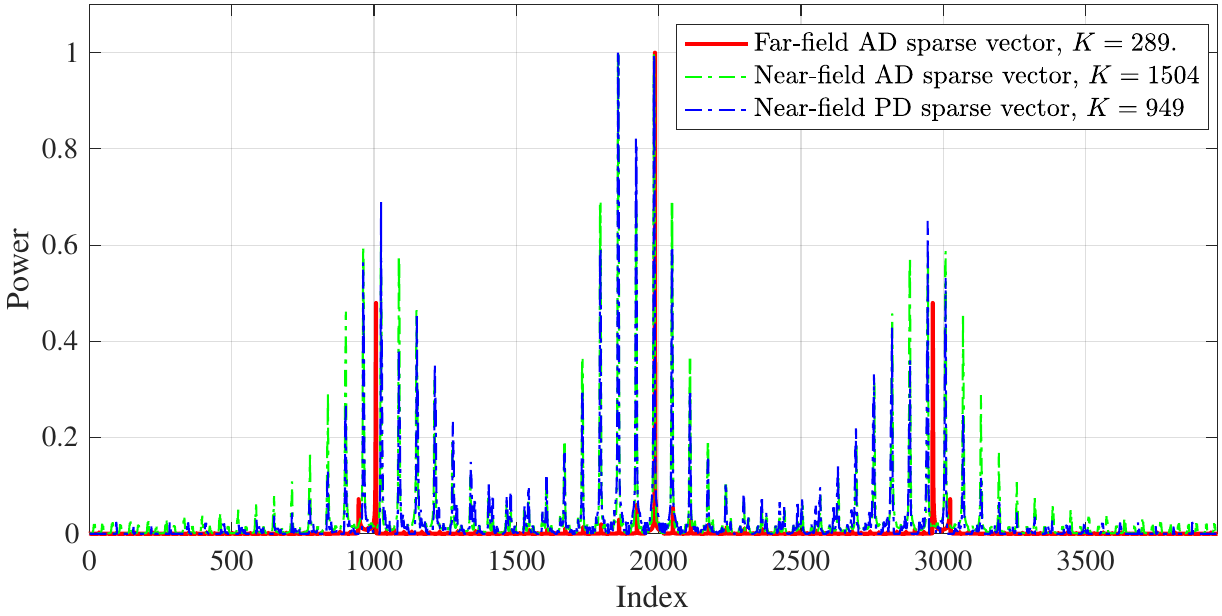}
		\caption{\color{black}Illustration of the weak sparseness in near-field communication with the ($64 \times 64$) UPA, number of scatterers is set to $L = 3$. $K$ denotes the number of non-zero entries.}
		\label{fig:blurred_supports}
	\end{minipage}
	\hfill 
	\begin{minipage}[b]{0.43\linewidth}
		\includegraphics[width=\linewidth]{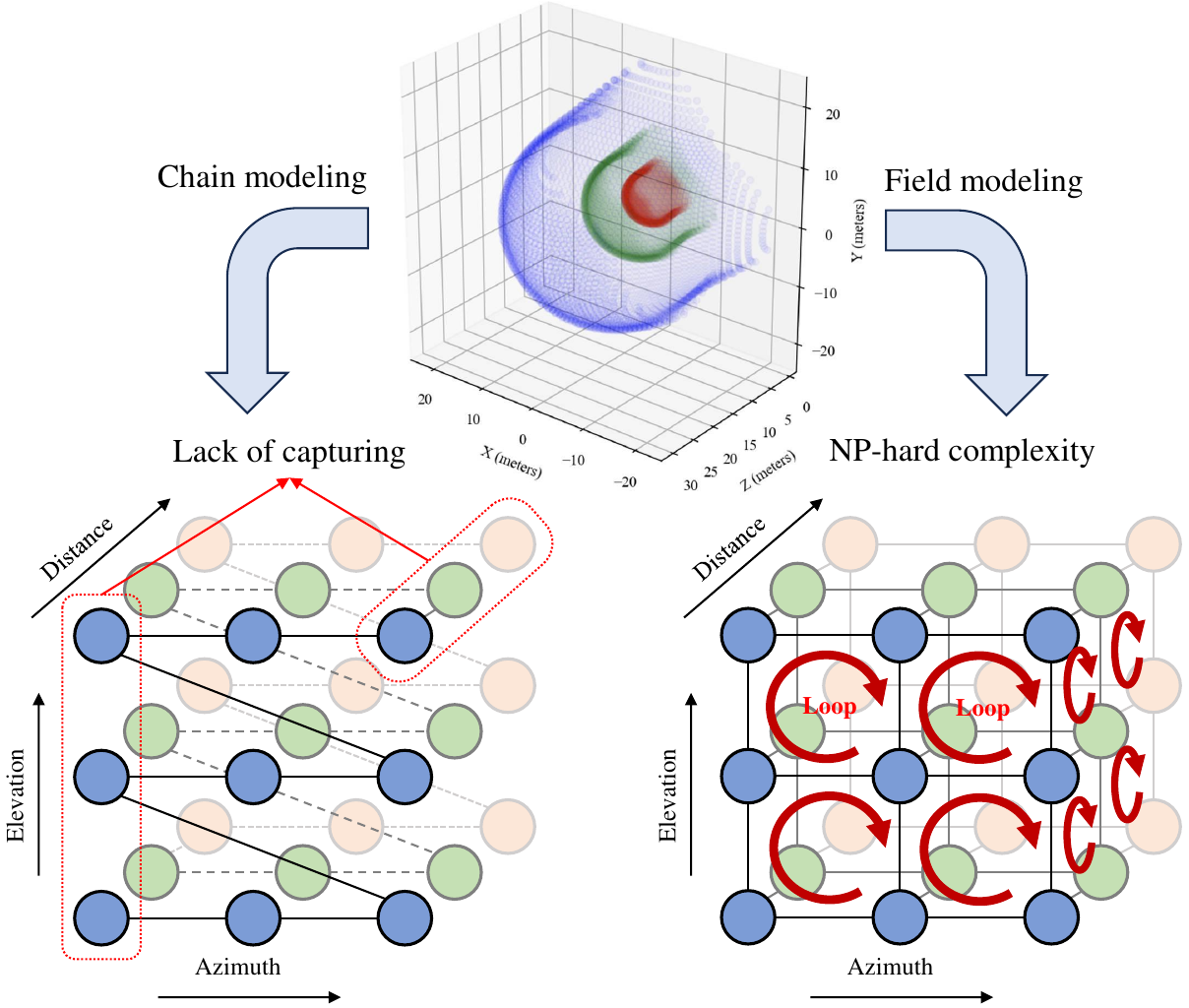}
		\caption{\color{black}Capturing structured information in 3D space.}
		\label{fig:3D_structure}
	\end{minipage}
  \vspace*{-0.5cm}
\end{figure*}

\subsection{
Mathematical Challenges of Applying Structured Sparsity in Near-Field Regime}
{\color{black}
In this section, we explore the mathematical challenges of applying structured sparsity in near-field communications with existing sparsifing methods. 
}


\subsubsection{\textbf{Weak Sparseness}}
  CS-based schemes rely significantly on the assumption of strong sparseness{\color{black}, where} in the presence of an appropriate {\color{black}dictionary basis}, a sparse vector can be attained with only a few non-zero entries included. In the far-field case, the channel can be straightforwardly represented to an AD sparse vector via the DFT basis~\cite{11.}. 
In Fig.~\ref{fig:blurred_supports}, we compare the sparse channel vectors in the far field using the AD sparsifying basis and the near-field counterparts using the AD and PD sparsifying basis, respectively.
The traditional AD sparse representation in the far field is characterized by a few non-zero entries, each of which corresponds to a significant path.
However, the AD basis fails to identify the significant channel power in the near field due to the power leakage effect. 
In particular, the non-zero entries no longer maintain a clear one-to-one mapping with significant angles, termed {\it weak sparseness}. This lack of clear mapping is attributed to the intricate couplings between angles and distances under the Fresnel approximation, leading to the AD power from a significant angle leaking to neighboring angles. Unfortunately, this issue presents a major obstacle in identifying significant paths within the near-field environment.

  To mitigate the weak sparseness, state-of-the-art efforts have proposed the PD basis~\cite{26., 29.}.
  Although the PD sparse representation leads to more concentrated and compact non-zero entries, the weak sparseness issue still exists.
  The main reason is that it can only assure approximate orthogonality between the basis vectors through a coherence controlling factor, which typically enlarges the basis coherence to 0.5~\cite{29.}. 
  This compromise on the orthogonality increases the ambiguity of dictionary basis, leading to high false detection probability, thereby offering no robust guarantee for sparse near-field channel recovery. 
    On the other hand, the design of the coherence controlling factor comes at the cost of sacrificing the size of the PD coverage region. This limitation on distance region becomes even more pronounced in uniform planar array (UPA) systems~\cite{26.}.

\subsubsection{\textbf{3D Structured Information}}
The concept of structured sparsity enhances the performance of CS-based methods by introducing additional {\it structured information} within the sparse representation. This newly introduced information lies in the correlation among the non-zero entries of the sparse channel representation. A host of strategies have delved into the HMM-based probabilistic frameworks to capture this structured information~\cite{11., 11.18, 11.18.1}. 
While chain-based models adeptly extract the structured information in 1D sparse vectors, field-based models are tailored for 2D sparse matrices. 
However, the triple-coupled structure of geometric parameters in near-field scenarios gives rise to an intractable 3D sparse cube, where neither chain nor field-based modeling can effectively encapsulate the near-field structured information. 
More precisely, as shown in Fig.~3, the use of chain models to represent near-field geometric parameters always neglects the structured information in the other two dimensions. While field models have the capacity to comprehensively capture structured information spanning all dimensions, they introduce computational hurdles, specifically loops burdened with NP-hard complexities. In summary, there still exists a notable gap in devising effective approaches specifically crafted to encapsulate the 3D structured information in the near-field regime.
Additionally, as illustrated in Fig.~3, the overall size of the 3D space is proportional to the product of the sizes of the elevation, azimuth, and distance dimensions, thereby contributing to the unacceptable multiplicative complexity.

\begin{figure*}
	\centering
	\includegraphics[width=.9\textwidth]{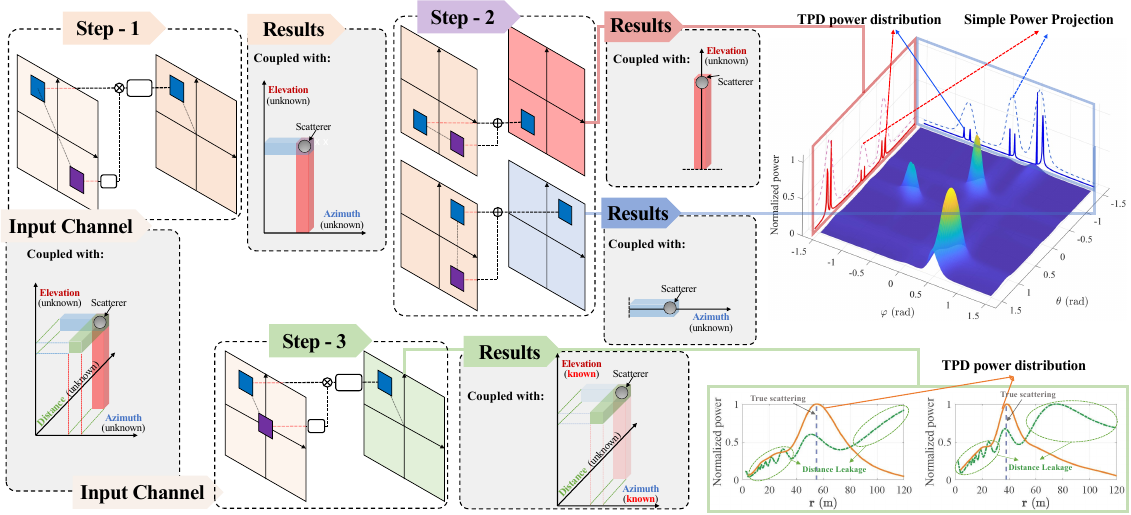}
	\caption{{\color{black}The proposed TPD framework for near-field communications.} }
	\label{fig:TPD_overlook}
	\vspace*{-0.5cm}
\end{figure*}



\section{\color{black}Triple Parametric Decomposition for Near-Field Channels}


    {\color{black}The crux of this pair of physical and mathematical problems stems from the coupling of the 3D geometric parameters under near-field conditions.} 
    {\color{black}Rather than additional sophisticated remedial measures addressing these challenges retroactively, we advocate for decoupling the 3D variables prior to the onset of these problems. 
    Guided by this rationale, we propose the TPD framework, the specifics of which will be elaborated in the subsequent section.}

\subsection{Implementation of the TPD Framework}
{\color{black}
    In near-field scenarios, the phase sequence at the antennas can be approximated by Fresnel expression, denoted by a quadratic polynomial, as opposed to the linear expressions commonly found in the far-field counterparts. 
    The non-linear Fresnel expression is the root source of the near-field peculiarities and complicates the application of structured sparsity in near-field systems. 
    Despite this, specific mathematical patterns can still be observed in the Fresnel quadratic polynomial. 
        To elaborate, the linear term in the polynomial exhibits a linearly monotonic property, while the quadratic term is both non-negative and symmetric w.r.t. the geometry center of the antenna array.
    In this context, the linear term captures the planar components within the near-field wavefront, containing only directional information related to the elevation-azimuth angle pair. Conversely, the quadratic term describes the curvature information in the spherical wavefront, encapsulating the distance information. 
This observation provides us with a handle for decoupling distance and angle pair variables, as shown in Fig.~\ref{fig:TPD_overlook}. 
}

\subsubsection{\color{black} \textbf{Step 1} -- Decomposition Between the Angle Pair and Distance}
{\color{black}
We employ a carefully designed strategy to enable the angular decomposition. By selecting a pair of antenna indices that are symmetrical to the origin, their expectation of the conjugate multiplication constitutes an entry of the angular-related channel response. In this case, the distance-related quadratic terms have been eliminated while the angle-pair-related linear terms are preserved. Consequently, we obtain a channel response that is exclusively related to the elevation-azimuth angle pair.}

\subsubsection{\color{black} \textbf{Step 2} -- Decomposition of Elevation-Azimuth Angle Pair}

{\color{black}To further decouple the elevation-azimuth angler pair, Step~2 employs a similar strategy to derive a channel response that only contains the elevation or azimuth angle. 
For each channel entry obtained in Step 1, we consider a pairing strategy that chooses the horizontally symmetrical entry to form a new channel entry pair, whose sum constitutes the elevation-related term in the phase and the azimuth-related term in the amplitude. 
Since the elevation angle estimation solely relies on the phase, leveraging the channel response derived in this manner for elevation angle estimation will no longer be influenced by the azimuth.
Similarly, due to the reciprocity of the elevation and azimuth angles, we can apply a vertically symmetrical channel entries pairing strategy, obtaining a channel response solely related to the azimuth angle.}
{\color{black}Differing from the simple projection-based approach, the proposed TPD can preserve the 2D structured information in the angle pair.}

\subsubsection{\color{black} \textbf{Step 3} -- Distance Extraction}
{\color{black}
    Although Step 1 successfully decoupled the distance variable from the elevation-azimuth angle pair, we only retained the angular information while completely overlooking the distance information. Consequently, retrieving the distance information from the original near-field channel is imperative. 
    Specifically, each channel entry is paired with the fixed entry at the UPA's geometric center. 
    Following the approach in Step 1, we conjugate-multiply the pair entries and take their expectations. Upon this operation, all the angle-related-linear and the distance-related-quadratic terms are fully retained. 
    Additionally, given that we have comprehensively captured the angle pair data in the sub-problems parsed in Step 2, and considering them as constant parameters, the problem is reduced to estimating the distances of scatterers within a 1D distance space.}

\begin{figure*}[t]
  \centering
    \begin{minipage}[b]{0.49\textwidth}
      \centering
      \includegraphics[width=0.98\linewidth]{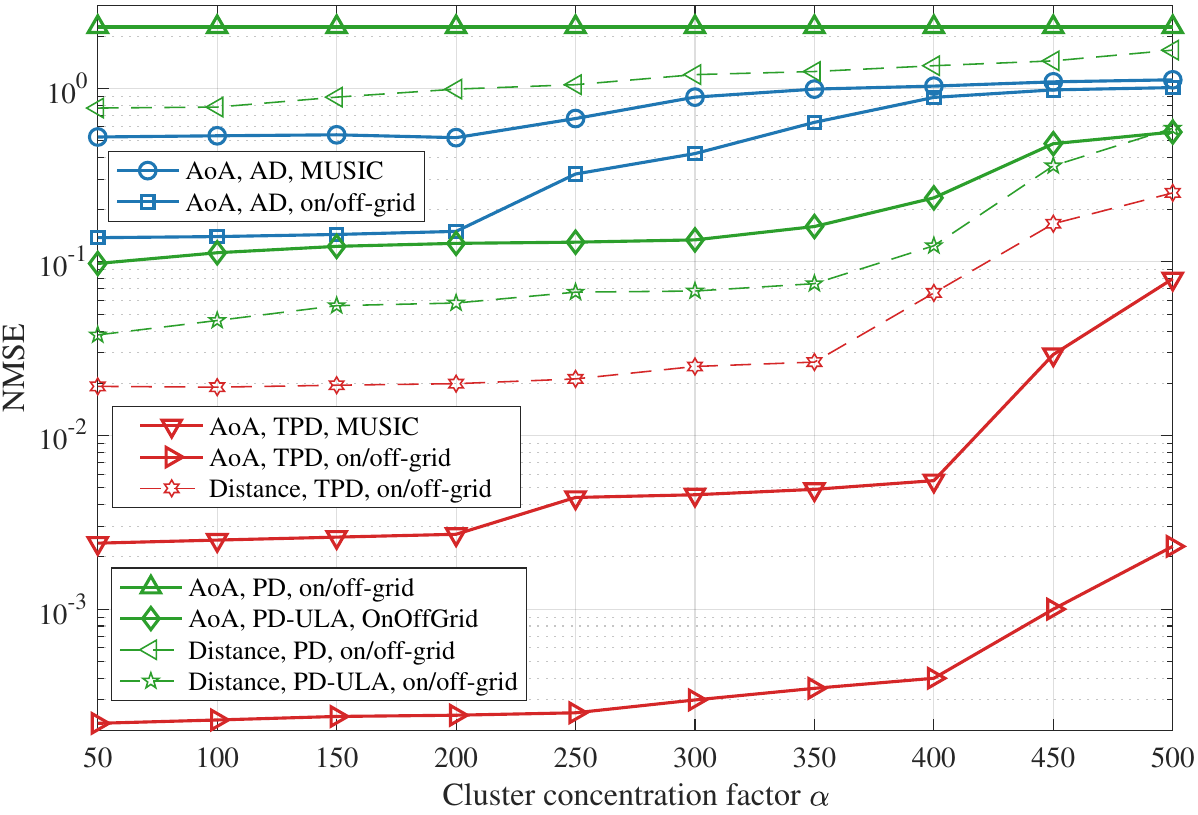}
      \captionsetup{justification=raggedright, singlelinecheck=false}
      \caption{\color{black}NMSE of geometric parameters versus cluster concentration factor, array size: ($256\times 256$ for default, $256\times 1$ for ULA setting), distance set to $10\ {\rm m}$, $3$ clusters, each of which contains $2$ scatterers.} 
      \label{fig:Simulation1}
    \end{minipage}
  \hfill
    \begin{minipage}[b]{0.49\textwidth}
      \centering
      \includegraphics[width=0.98\linewidth]{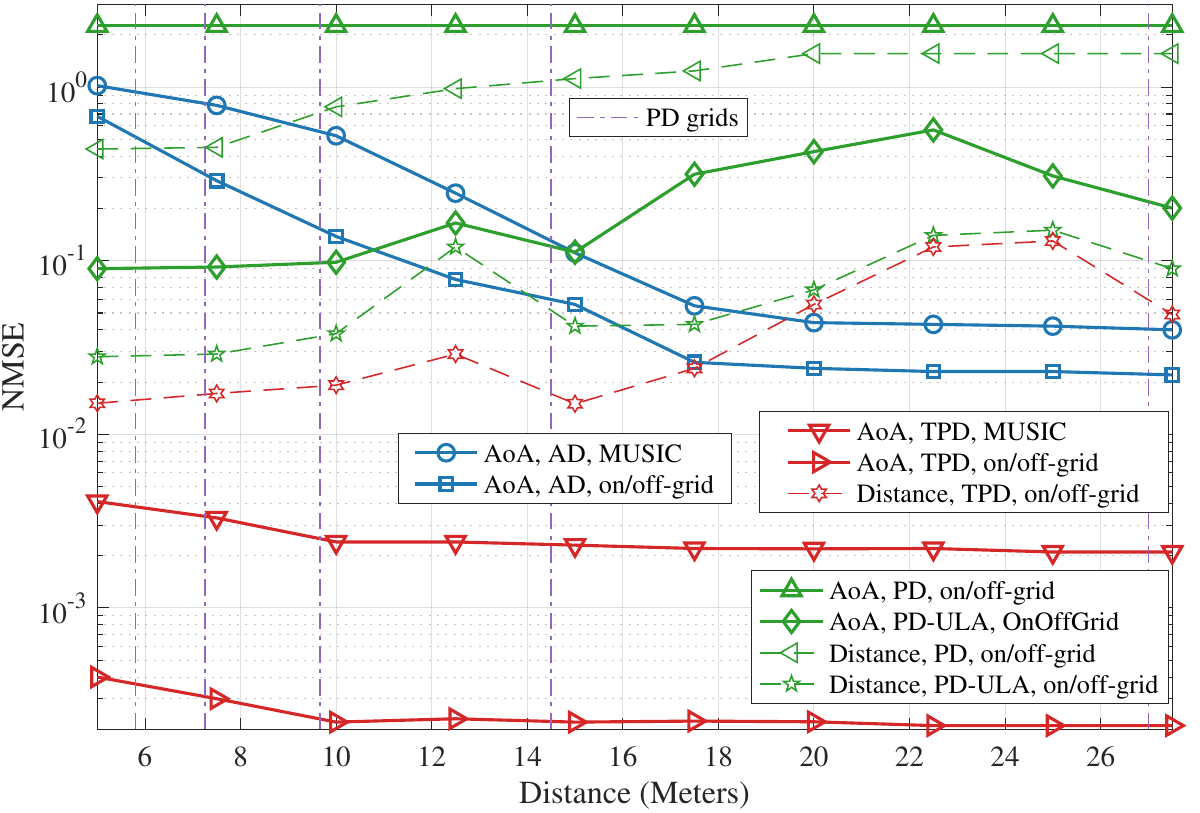}
      \captionsetup{justification=raggedright, singlelinecheck=false}
      \caption{\color{black}NMSE of geometric parameters versus distance between scatterers and the array, cluster concentration factor set to $50$, Rayleigh and Fresnel distance are $4.6\ {\rm m}$ and $175.5\ {\rm m}$, respectively.} 
      \label{fig:Simulation2}
    \end{minipage}
  \vspace*{-0.5cm}
\end{figure*}


\subsection{\color{black} Advantages of the TPD Framework}
{\color{black}
We compared the performance of TPD with other benchmarks in estimating geometric parameters under the influence of cluster concentration factor and the distance between scatterers and the array, as illustrated in Fig.~\ref{fig:Simulation1} and Fig.~\ref{fig:Simulation2}, respectively. Specifically, the cluster concentration factor of the von Mises-Fisher (vMF)-based channel model controls the cluster size~\cite{28.6.}, a higher value of which results in a smaller cluster size.
Since TPD provides sparsifying method for the near-field channel, we compare the performance of TPD with the widely-used angular-domain (AD) sparsifying method~\cite{11.,22.,12.} and the recently proposed polar-domain (PD) sparsifying method both in UPA and uniform linear array (ULA) setting~\cite{26., 29.}. 
Furthermore, to demonstrate the comparability of TPD with the underlying algorithm, we not only used the state-of-the-art on/off-grid compressive sensing algorithm under various sparsifying methods but also compared the performance when utilizing the traditional multiple signal classification (MUSIC) algorithm~\cite{Spectral_CS}.
}

\subsubsection{\color{black} \textbf{Robustness Against Cluster Size}}
{\color{black}The reduction in cluster size leads to an excessively small angular interval among scatterers, making them difficult to distinguish. In far-field MIMO channels, this issue can be effectively addressed by enlarging the MIMO array size, thus ensuring the angular-domain resolution is sufficient to discern the angular intervals of scatterers. However, in near-field channels, the problem of power leakage results in the overlapping of spreading spectra, which complicates the extraction of sparsity in the near-field. Consequently, the performance of the angular-domain method suffers fast degradation as the cluster concentration factor increases. While the polar-domain method can achieve robustness comparable to the proposed TPD in ULA scenarios, its excessive correlation among sparsifying bases leads to a high incidence of false detections of scatterers in UPA scenarios. In contrast, the TPD framework proposed in this study demonstrates markedly superior accuracy and robustness over traditional sparsifying methods.}

\subsubsection{\color{black} \textbf{Robustness Against Distance}}
{\color{black}Generally, within near-field channels, the proximity of scatterers to the array significantly exacerbates the effects of power leakage brought about by spherical waves. Conversely, the channel increasingly resembles a far-field channel model as the distance increases. Thus, the performance of traditional angular-domain-based methods deteriorates noticeably as the distance decreases. In contrast, the efficacy of polar-domain-based methods diminishes with increasing distance. This decline in performance is attributable to the reliance of the polar-domain basis on the distance grid distribution. Given the inverse-ratio-based design of distance grids~\cite{29.}, greater distances result in sparser grid distribution, leading to decreased performance.
Distinct from both angular- and polar-domain methods, after the process in Step 1 of the TPD, the elevation-azimuth angle pair is decoupled from the distance variable. In simpler terms, TPD isolates the planar wave component from the distance-associated spherical wavefront, significantly alleviating the power leakage problem and thereby enhancing robustness against the distance variable.}

\subsubsection{\color{black} \textbf{Broad-Spectrum Compatibility}}
{\color{black} 
The proposed TPD framework operates at a higher level, offering a decoupled sparse channel representation. Hence, it facilitates integration across a variety of algorithms. For instance, as demonstrated in Fig.~\ref{fig:Simulation1} and Fig.~\ref{fig:Simulation2}, the TPD framework significantly enhances the MUSIC algorithm, outperforming methods in both the angular and polar domains.
    The TPD framework can be successfully implemented once the antenna array possesses central symmetry and horizontal/vertical symmetry. Consequently, beyond its compatibility across different algorithms, it also boasts extensive adaptability across various scenarios.
}

\subsubsection{\color{black} \textbf{Reduced Structural and Computational Complexity}}
{\color{black}The sparse representation facilitated by the TPD is divided into three distinct vectors, corresponding solely to elevation, azimuth, and distance, respectively. This allows for the individual capture of structured information inherent in these geometric parameters, obviating the need for more intricate models customized for a triply-coupled 3D structure. Consequently, this approach reduces the structural complexity inherent in algorithm design and allows for more flexible design.
On the other hand, the dimension of the coupled 3D geometric parameters results from multiplying the dimensions of elevation, azimuth, and distance.  If sparse recovery is directly carried out in the near-field case, the algorithm must search for the 3D geometric parameters with a solution space of $\mathcal{O}(N^{2.5})$ in an $(N\times N)$-antenna near-field system \cite{TVT}. By contrast, within the proposed TPD framework, the geometric parameters shift from being contained within a 3D coupled cube to being represented by three independent vectors. This leads to a significant reduction in dimensional complexity: transitioning from the multiplicative $\mathcal{O}(N^{2.5})$ to additive $\mathcal{O}(2.5N)$.}

\section{\color{black}Applications and Future Directions of the TPD Framework}

In this section, we explore the prospective applications alone with the future directions of the TPD framework for near-field communications.

\subsection{\color{black}TPD Applications Enhanced by Structured Sparsity}

\subsubsection{Clustered Sparsity in Robust AoA Detection}
A typical kind of structured sparsity in near-field channels is the clustering scatterers in wireless propagation environments~\cite{TVT}. 
This results in grouped non-zeros elements in the sparse representations.
By applying the TPD process to address power leakage issues, the imposed sparsity is attained in the decoupled sparse vector for each dimension, providing more precise cluster structure representations to the underlying CS-based algorithms. 

\subsubsection{{Temporal} Sparsity in Recursive Channel Tracking}
{\color{black}
    Besides the spatial structured sparsity such as clustered sparsity, structured information in the temporal domain also exists, termed temporal sparsity. 
    Specifically, in recursive channel tracking problems, the scatterers from the previous time slot offer prior information for the current time slot.
    {\color{black}
        However, within the near-field 3D geometric parameters, the structured information across time slots is hidden in the triple-coupled structure. 
        Specifically, the state transition function is the joint function w.r.t. all the geometric dimensions.
        {\color{black}
            This leads to exceeding computational complexity and strong correlation-induced cross-dimension interference.}
        Fortunately, The original joint 3D parametric tracking problem can be simplified through the TPD filtering to three distinct parametric tracking problems with a single DoF.}
    As such, even the simplest chain-based model can be used to provide sufficient temporal information to enhance tracking accuracy.}
    
\subsubsection{\color{black} Common Sparsity in location division multiple access (LDMA)}
    The distance ingredient can be employed for enhancing multiple access techniques by introducing a new DoF, i.e., LDMA~\cite{26.}.
    {\color{black}
        The scatterers within each user's distinct near-field channel {\color{black}may share the same locations on the elevation and azimuth dimensions}, referred to as {\it common sparsity}~\cite{12.}.}
    {\color{black}
        Specifically, different users may share the same geometric parameters distribution on one dimension, despite the fact that they are located in different 3D geometric positions.}
    {\color{black}
        We can design the underlying algorithm to search the common angles (or distances) first and then search for the distinct distances across users.
        Therefore, the searching space can be compressed from multiplicative to additive, paving the way for low-complexity and low-overhead LDMA designs.}
    }

\subsection{Future Directions}



\subsubsection{Structured Sparsity in Continuous Aperture Design}
{\color{black}Future holographic MIMO (HMIMO) surface aims to achieve unprecedentedly finer beamforming in free space to attain unparalleled spacial multiplexing gains~\cite{8., 28.6.}. 
This will be achieved by packing nearly infinite meta-surface-based entries on the array plane with the antenna spacing far less than half the wavelength. 
This continuous or near-continuous aperture design will profoundly revolutionize electromagnetic channel modeling.
Similarly, the proposed TPD framework can also be effectively implemented in HMIMO systems. 
{
  Specifically, we can redesign the dictionary based on the Fourier harmonics~\cite{28.6.}, where each dictionary entry denotes a corresponding Fourier-harmonic-based wavenumber-domain (WD) element.
  The reason is that the continuous aperture design will lead to a near-infinite number of traditional DFT basis, while the number of WD basis is solely determined by the ratio of the antenna aperture and the working frequency.}
By performing the TPD framework on the Fourier harmonics and sparsifying the channel in the corresponding wavenumber domain, we can extend the advantages of the TPD framework from traditional discrete MIMO to HMIMO systems with continuous antenna aperture design.}

\subsubsection{Efficient Algorithm Design for Sparse Recovery}
Although the TPD framework can alleviate issues such as power leakage and weak sparsity, CS-based methods inherently possess quantization errors, acting as performance bottlenecks for sparse channel recovery.  
Specifically, offsets exist between the discrete non-zero entries in the sparse channel representations and the real geometric parameters~\cite{11.}.
Hence, a two-module algorithm design is needed to improve sparse recovery performance in the TPD framework, where the first aims to achieve sparse signal recovery in the discrete sparse domain; the second module, using techniques such as successive convex optimization and gradient descent, focuses on precisely estimating the quantization errors. 
With this dual-module design, we can further improve the performance of the CS-based algorithm with the assistance of structured sparsity.

\section{Conclusion}
{\color{black}
This article investigates the challenges, peculiarities, and applications from the perspective of the near-field structured sparsity.
In particular, by revisiting the prior works devoted to the structured sparsity in the far field, we elucidate the fundamental physical peculiarities induced by the tightly coupled 3D parameters in the near field.
Upon this, we expose the consequent mathematical challenges by detailing the theoretical limitations of various conventional sparsifying methods when attempting to extract near-field structured sparsity.
Guided by the rationale of decoupling, a low-complexity TPD framework is proposed to decompose triple-coupled geometric parameters, facilitating their individual sparse recovery.
Then, we outline several applications in which the TPD framework can effectively be employed to realize parametric recovery based on decoupled structured sparsity.
We finally envisage promising directions for future research within this domain.}

\vspace{-0.2cm}
\bibliographystyle{IEEEtran}
\bibliography{CM-TPD}                       






\end{document}